\documentclass[conference]{IEEEtran}
\usepackage{graphicx,amsmath,latexsym,amssymb}
\ifCLASSINFOpdf
\else
\fi
\begin{document}

\title{Spectral Efficiency of MIMO Millimeter-Wave Links with Single-Carrier Modulation
for 5G Networks}

\author{\IEEEauthorblockN{Stefano Buzzi, Carmen D'Andrea}
\IEEEauthorblockA{DIEI - Universit\`a di Cassino  \\ e del Lazio Meridionale\\
I-03043 Cassino (FR) - Italy}
\and
\IEEEauthorblockN{Tommaso Foggi, Alessandro Ugolini, \\  Giulio Colavolpe}
\IEEEauthorblockA{DII - Universit\`a di Parma\\
I-43124, Parma - Italy}
}


\maketitle

\begin{abstract}
Future wireless networks will extensively rely upon bandwidths 
centered on carrier frequencies larger than 10GHz. Indeed, recent research has shown that, despite the large path-loss, millimeter wave (mmWave) frequencies can be successfully exploited to transmit  very large data-rates over short distances to slowly moving users. Due to hardware complexity and cost constraints, single-carrier modulation schemes, as opposed to the popular multi-carrier schemes, are being considered for use at mmWave frequencies.  
This paper presents preliminary studies on the achievable spectral efficiency on a wireless MIMO link operating at mmWave in a typical 5G scenario. Two different single-carrier modem schemes are considered, i.e. a traditional modulation scheme with linear equalization at the receiver, and a single-carrier modulation with cyclic prefix, frequency-domain equalization and  FFT-based  processing
at the receiver.
Our results show that  the former achieves a larger spectral efficiency than the latter. Results also confirm that the spectral efficiency increases with the dimension of the  antenna array, as well as that performance gets severely degraded when the link length exceeds 100 meters and the transmit power falls below 0dBW.
Nonetheless, mmWave appear to be very suited for providing very large data-rates over short distances.

\end{abstract}


%
\IEEEpeerreviewmaketitle

\section{Introduction}

The research on the next generation of wireless networks is proceeding at an intense pace, both in industry and in academia. Focusing on the Physical Layer, there is wide agreement \cite{whatwillbe} that fifth-generation (5G) wireless networks will be based, among the others, on three main innovations with respect to legacy fourth-generation systems, and in particular (a) the use of large scale antenna arrays, a.k.a. massive MIMO \cite{massiveMIMO}; (b) the use of small-size cells in areas with very large data request \cite{smallcells}; and (c) the use of carrier frequencies larger than 10GHz \cite{6515173}. 

Indeed, focusing on (c), the use of the so-called millimeter wave (mmWave) frequencies has been proposed as a strong candidate approach to achieve the spectral efficiency growth required by 5G wireless networks, resorting to the use of currently unused frequency bands in the range between $20\,\textrm{GHz}$ and $90\,\textrm{GHz}$. In particular, the E-band between $70\,\textrm{GHz}$ and $80\,\textrm{GHz}$ provides $10\,\textrm{GHz}$ of free spectrum which could be exploited to operate 5G networks. It is worth underlining that mmWave are not intended to replace the use of lower carrier frequencies traditionally used for cellular communications, but rather as additional frequencies that can be used in densely crowded areas for short-range communications. 
Until now, the use of mmWave for cellular communications has been neglected due to the higher atmospheric absorption that they suffer compared to other frequency bands and to the larger values of the free-space path-loss. However, recent measurements suggest that mmWave attenuation is only slightly worse than in other bands, as far as propagation in dense urban environments and over short distances (up to about 100 meters) is concerned \cite{mmWaverecent}. Additionally, since antennas at these wavelengths are very small,  arrays with several elements can be packed in small volumes, in principle also on mobile devices, thus removing the traditional constraint that only few antennas can be placed on a smartphone and benefiting of an array gain at both edges of the communication link with respect to traditional cellular links. 
Another peculiar feature of cellular communications at mmWave that has been found is that these are mainly noise-limited and not interference-limited systems, and this will simplify the implementation of interference-management and resource-scheduling policies.
Based on this encouraging premises, a large body of work has been recently carried out on the use of mmWave for cellular communications \cite{6515173,mmWaverecent,RappaportGutierrezBen-DorMurdockQiaoTamir2013,coverageandcapacitymmWave, hybridprecodingmmwaves}.

One of the key questions about the use of mmWave is about the type of modulation that will be used at these frequencies. Indeed, while it is not even sure that 5G systems will use orthogonal frequency division multiplexing (OFDM) modulation at classical cellular frequencies \cite{BaBuCoMoRuUg14}, there are several reasons that push for 5G networks operating a single-carrier modulation (SCM) at mmWave \cite{mmWaverecent}. First of all, the propagation attenuation of mmWave make them a viable technology only for small-cell, dense networks, where few users will be associated to any given base station, thus implying that the efficient frequency-multiplexing features of OFDM may not be really needed. Additionally, 
the large bandwidth would cause low OFDM symbol duration, which, coupled with small propagation delays, means that the users may be multiplexed in the time domain as efficiently as in the frequency domain. Finally, 
 mmWave will be operated together with massive antenna arrays to overcome propagation attenuation. This makes digital beamforming unfeasible, since the energy required for digital-to-analog and analog-to-digital conversion would be huge. Thus, each user will have an own radio-frequency beamforming, which requires users to be separated in time rather than frequency.

\begin{figure}[!h]
\centering
\includegraphics[scale=0.34]{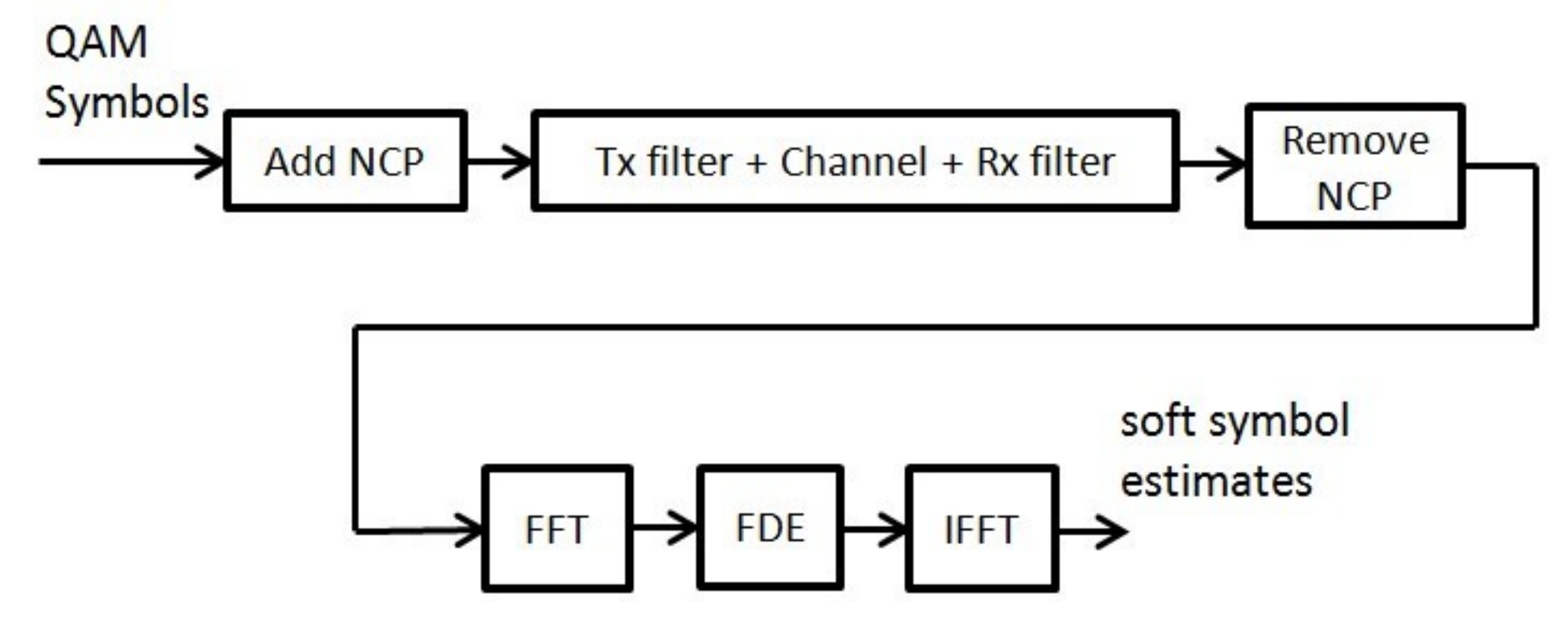}
\caption{Principle of NCP-SCM transceiver architecture; FDE stands for "frequency-domain equalization".}
\label{Fig:CP-SCM}
\end{figure}

In light of these considerations, SCM formats are being seriously considered for mmWave systems. 
For efficient removal of the intersymbol interference induced by the frequency-selective nature of the channel, the use of SCM coupled with a cyclic prefix has been proposed, so that FFT-based processing might be performed at the receiver \cite{CP-SC}
In \cite{Cudak13,Cudak2}, the null cyclic prefix single carrier (NCP-SC) scheme has been proposed for mmWave. The concept is to transmit a single-carrier signal, in which the usual cyclic prefix used by OFDM is replaced by nulls appended at the end of each transmit symbol. The block scheme is reported in Fig. \ref{Fig:CP-SCM}.

This paper is concerned with the evaluation of the achievable spectral efficiency (ASE) of SCM schemes operating over MIMO links at mmWave frequencies. We consider two possible transceiver architectures: (a) SCM with linear minimum mean square error (LMMSE) equalization in the time domain for intersymbol interference removal and symbol-by-symbol detection; and (b) SCM with cyclic prefix and FFT-based processing and LMMSE equalization in the frequency domain at the receiver. 
By adopting, inspired by \cite{spatiallysparse_heath,heath_selective}, a modified statistical MIMO channel model for mmWave frequencies, and using the simulation-based technique for computing information-rates reported in \cite{ArLoVoKaZe06}, we thus provide a preliminary assessment of the achievable spectral efficiency (ASE) that can be reasonably expected in a scenario representative of a 5G environment.  Our results show that, for distances less than 100 meters, and with a transmit power around 0dBW, mmWave links exhibit good performance and may provide good spectral efficiency; for larger distances instead, either large values of the transmit power or 
a large number of antennas must be employed to overcome the distance-dependent increased attenuation. 

The rest of this paper is organized as follows. Next Section contains the system model, with details on the two considered transceiver architectures and on the pulse shapes considered in the paper. Section III explains the used technique for the evaluation of the ASE, while extensive numerical results are illustrated and discussed in Section IV. Finally, Section V contains concluding remarks.


\begin{figure*}[t]
\centering
\includegraphics[scale=0.35]{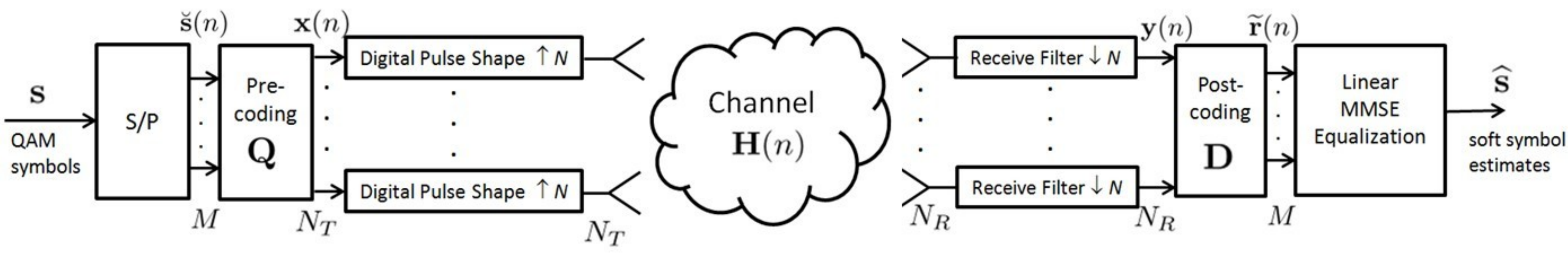}
\caption{Transceiver architecture with time-domain equalization.}
\label{fig:scenario1}
\end{figure*}

\section{System model}
We consider a transmitter-receiver pair that may be representative of either the uplink or the downlink of a cellular system. We denote by $N_T$ and $N_R$ the number of transmit and receive antennas, respectively.
Denote by $\mathbf{s}$ a column vector containing the $L$ data-symbols (drawn from a QAM constellation with average energy $P_T$)  to be transmitted:
\begin{equation}
\mathbf{s}=[s_0, s_1, \ldots, s_{L-1}]^T \; ,
\end{equation}
with $(\cdot)^T$ denoting transpose.
We assume that $L=kM$, where $k$ is an integer and  $M$ is the number of information symbols that are 
simultaneously transmitted by the $N_T$ transmit antennas in each symbol interval. 
The propagation channel is modeled in discrete-time as a matrix-valued finite-impulse-response (FIR) filter; in particular, we denote by $\left\{\mathbf{H}(n)\right\}_{n=0}^{P-1}$ the sequence, of length $P$, of the  $(N_R \times N_T)$-dimensional matrices describing the channel. The discrete-time versions of the impulse response of the transmit and receive shaping filters are denoted as $h_{TX}(n)$ and  
$h_{RX}(n)$, respectively; these filters are assumed to be both of length $P_h$.

We focus on two different transceiver architectures, one that operates equalization in the time-domain and one that works in the frequency domain through the use of a cyclic prefix.

\subsection{Transceiver model with time-domain equalization (TDE)}
We refer to the discrete-time block-scheme reported in Fig. \ref{fig:scenario1}. 
The QAM symbols in the vector $\mathbf{s}$ are fed to a serial-to-parallel conversion block that splits them in $k$ distinct $M$-dimensional vectors $\breve{\mathbf{s}}(1) , \ldots, \breve{\mathbf{s}}(k) $. These vectors are pre-coded using the 
the $(N_T \times M)$-dimensional precoding matrix $\mathbf{Q}$; we thus obtain the $N_T$-dimensional vectors 
$$
\mathbf{x}(n)=\mathbf{Q}\breve{\mathbf{s}}(n) \; , \qquad n=1, \ldots, k \; .
$$
The vectors $\mathbf{x}(n)$ are fed to a bank of $N_T$ identical shaping filters, converted to RF and transmitted. 

At the receiver, after baseband-conversion, the $N_R$ received signals are passed through a bank of filters matched to the ones used for transmission and sampled at symbol-rate. We thus obtain the $N_R$-dimensional vectors $\mathbf{y}(n)$, which are passed through a post-coding matrix, that we denote by  $\mathbf{D}$,  of dimension $(N_R \times M)$. 
Denoting by $\widetilde{\mathbf{H}}(n)$ the matrix-valued FIR filter representing the composite channel impulse response (i.e., the convolution of the transmit filter, actual matrix-valued channel and receive filter), it is easy to show that the generic $M$-dimensional vector at the output of the post-coding matrix, say $\widetilde{\mathbf{r}}(n)$,  is written as
\begin{equation}
\widetilde{\mathbf{r}}(n)= \mathbf{D}^H \mathbf{y}(n)=\displaystyle \sum_{\ell=0}^{\widetilde{P}-1}
\mathbf{D}^H \widetilde{\mathbf{H}}(\ell)\mathbf{Q}\breve{\mathbf{s}}(n-\ell) + \mathbf{D}^H\mathbf{w}(n) \; ,
\label{eq:received_signal_1}
\end{equation}
with $(\cdot)^H$ denoting conjugate transpose.
In (\ref{eq:received_signal_1}), $\widetilde{P}=P+2P_h-1$ is the length of the matrix-valued composite channel impulse response 
$\widetilde{\mathbf{H}}(n) $, while $\mathbf{w}(n)$ is the additive Gaussian-distributed thermal noise at the output of the reception filter. 
Regarding the choice of the pre-coding and post-coding matrices $\mathbf{Q}$ and $\mathbf{D}$, 
letting $\eta=\mbox{arg}\max_{\ell = 0, \ldots, \widetilde{P}-1}\left\{ \left\|\widetilde{\mathbf{H}}(\ell)\right\|_{\rm F}\right\}$, with $\| \cdot \|_{\rm F}$ denoting the Frobenius norm, 
we assume here that $\mathbf{Q}$ contains on its columns the left eigenvectors of the matrix $\widetilde{\mathbf{H}}(\eta)$ corresponding to the $M$ largest eigenvalues, and that the matrix $\mathbf{D}$ contains on its columns the corresponding right eigenvectors\footnote{Note that, due to the presence of intersymbol interference, the proposed pre-coding and post-coding structures are not optimal. Nevertheless, we make here this choice for the sake of simplicity. The proposed pre-coding and post-coding structures are also fully digital; the design of hybrid, i.e. mixed analog-digital structures, and the evaluation of the corresponding ASE is an interesting issue left for future work.}. 

In order to combat the intersymbol interference, an LMMSE equalizer is used. In particular, to obtain a soft estimate of the data vector $\breve{\mathbf{s}}(n)$, the $\widetilde{P}$ observables $\widetilde{\mathbf{r}}(n), \widetilde{\mathbf{r}}(n+1), \ldots, 
\widetilde{\mathbf{r}}(n+\widetilde{P}-1)$ are stacked into a single $\widetilde{P}M$-dimensional vector, that we denote by 
$\widetilde{\mathbf{r}}_{\widetilde{P}}(n)$, and processed as follows:
\begin{equation}
\widehat{\breve{\mathbf{s}}}(n)=\mathbf{E}^H\widetilde{\mathbf{r}}_{\widetilde{P}}(n) \; ,
\label{eq:processingTDE}
\end{equation}
where $\mathbf{E}$ is a $[\widetilde{P}M \times M]$-dimensional matrix representing the LMMSE equalizer\footnote{We do not report here its explicit expression for the sake of brevity. A good reference about LMMSE estimation is the textbook \cite{kay1998}.}.

\noindent
{\em Considerations on complexity.} Regarding processing complexity, we note that the computation of the equalization matrix $\mathbf{E}$ requires the inversion of the covariance matrix of the vector 
$\widetilde{\mathbf{r}}_{\widetilde{P}}(n)$,  with a computational burden proportional to $(\widetilde{P}M)^3$;
then,  implementing Eq. \eqref{eq:processingTDE} requires a matrix vector product, with a computational burden proportional to $(\widetilde{P} M^2)$; this latter task must be made $k$ times in order to provide the soft vector estimates for all values of $n=1, \ldots, k.$

\begin{figure*}[!t]
\centering
\includegraphics[scale=0.35]{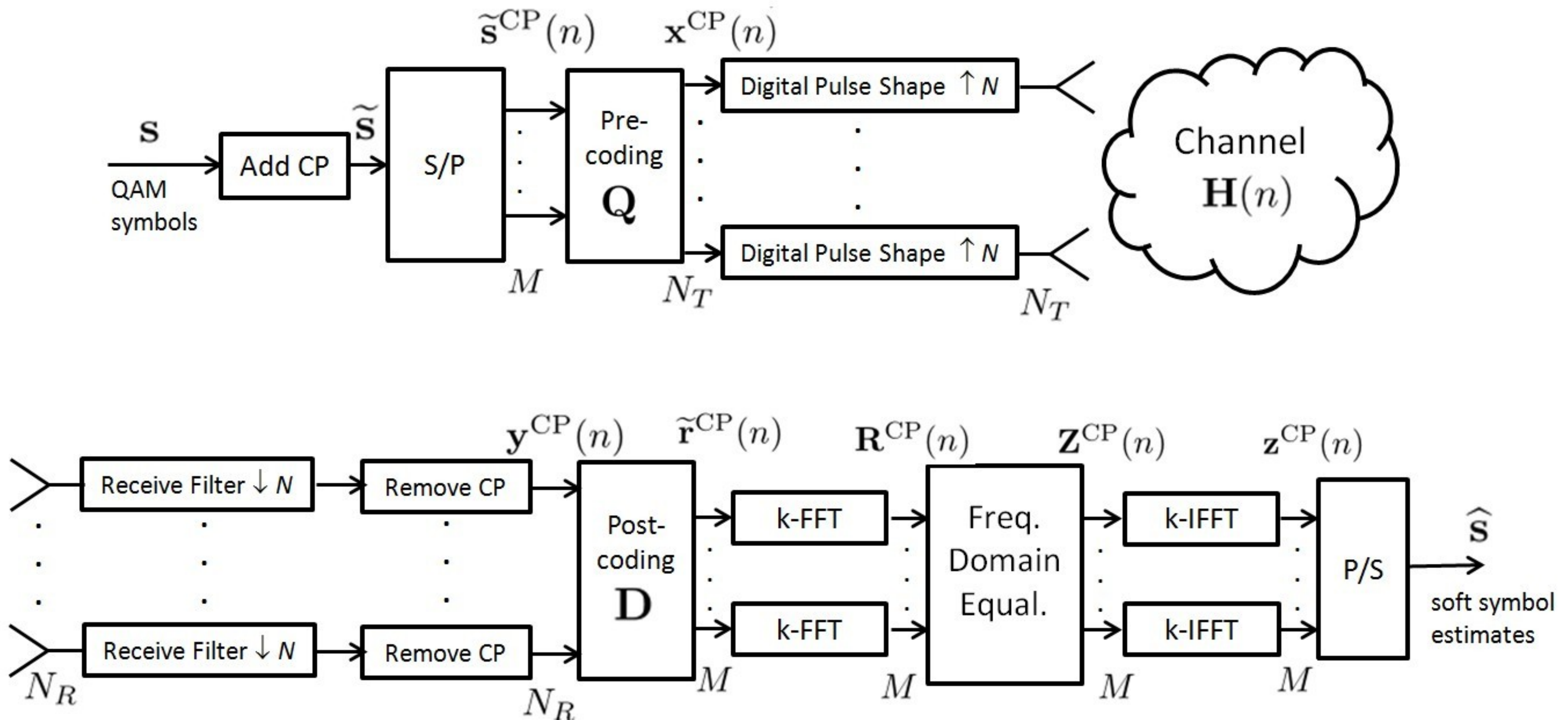}
\caption{Transceiver architecture with cyclic prefix, FFT-based processing and frequency-domain equalization.}
\label{fig:scenario2}
\end{figure*}

\subsection{Transceiver model with frequency-domain equalization (FDE)}
We refer to the discrete-time block-scheme reported in Fig. \ref{fig:scenario2}. 
A CP of length $CM$ is added at the beginning of the block $\mathbf{s}$ of $L=kM$ QAM symbols, so as to have the vector $\widetilde{\mathbf{s}}$ of lenght $(k+C)M$.
As in the previous case, the vector $\widetilde{\mathbf{s}}$ is passed through a serial-to-parallel conversion with $M$ outputs, a precoding block (again expressed through the matrix $\mathbf{Q}$), a bank of $N_T$ transmit filters; then conversion to RF and transmission take place.
At the receiver, after baseband-conversion, the $N_R$ received signals are passed through a bank of filters matched to the ones used for transmission and sampled at symbol-rate; then, the cyclic prefix is removed.  We thus obtain the $N_R$-dimensional vectors
$\mathbf{y}^{\rm CP}(n)$, with $n=1, \ldots, k$, containing a noisy version of the \textit{circular} convolution between the sequence $\mathbf{x}^{\rm CP}(n)$ and $\widetilde{\mathbf{H}}(n)$ , i.e.:
\begin{equation}
\mathbf{y}^{\rm CP}(n)=
 \widetilde{\mathbf{H}}(n) \circledast \mathbf{x}^{\rm CP}(n)  + \mathbf{w}(n) \; , \qquad n=1, \ldots, k
\end{equation}
The vectors $\mathbf{y}^{\rm CP}(n)$ then 
are processed by the post-coding matrix $\mathbf{D}$ (the choice of the matrices $\mathbf{Q}$ and $\mathbf{D}$ is the same as in the TDE case, so it is not repeated here); we thus obtain the $M$-dimensional vectors $\widetilde{\mathbf{r}}^{\rm CP}(n)=\mathbf{D}^H \mathbf{y}^{\rm CP}(n)$, with $n=1, \ldots, k$.
These vectors 
go through an entry-wise FFT transformation on $k$ points; the $n$-th FFT coefficient, with $n=1, \ldots, k$, can be shown to be expressed as
\begin{equation}
\mathbf{R}^{\rm CP}(n)= \widetilde{{\cal H}}(n) \mathbf{X}^{\rm CP}(n) + \mathbf{W}(n) \; ,
\label{eq:fftdomain}
\end{equation}
where $\widetilde{{\cal H}}(n)$ is an $(M \times N_T)$-dimensional matrix representing the $n$-th FFT coefficient of the matrix-valued sequence $\mathbf{D}^H \widetilde{\mathbf{H}}(n)$, and $ \mathbf{X}^{\rm CP}(n)$ and $\mathbf{W}(n)$ are the $n$-th FFT coefficient of the sequences $\mathbf{x}^{\rm CP}(n)$ and $\mathbf{D}^H \mathbf{w}(n)$, respectively.
From Eq. (\ref{eq:fftdomain}) it is seen that, due to the presence of multiple antennas, and, thus, of the matrix-valued channel, the useful symbols reciprocally interfere and thus an equalizer is needed.
We denote by $\mathbf{E}(n)$ the $(M \times M)$-dimensional equalization matrix; a zero-forcing approach can be adopted here, i.e. we let  $\mathbf{E}^H(n) =  (\widetilde{{\cal H}}(n) \mathbf{Q})^{-1}$, and the output of the equalizer can be shown to be written as
$$
\mathbf{Z}^{\rm CP}(n)= \mathbf{E}^H(n) \mathbf{R}^{\rm CP}(n)=\widetilde{\mathbf{S}}^{\rm CP}(n) +  
 (\widetilde{{\cal H}}(n) \mathbf{Q})^{-1}\mathbf{W}(n) \; .
$$
In the above equation, $\widetilde{\mathbf{S}}^{\rm CP}(n)$ is an $M$-dimensional vector representing the $n$-th FFT coefficient of the vector-valued sequence $\widetilde{\mathbf{s}}^{\rm CP}(n)$ -- we are using here the equation 
$\mathbf{X}^{\rm CP}(n)= \mathbf{Q} \widetilde{\mathbf{S}}^{\rm CP}(n)$, which can be shown with ordinary efforts.

Then, the vectors 
$\mathbf{Z}^{\rm CP}(n)$ go through an entry-wise IFFT transformation on $k$ points, which yields the soft symbol estimates of the entries of the data vector $\mathbf{s}$.

\noindent
{\em Considerations on complexity.} Looking at the scheme in Fig. \ref{fig:scenario2}, the computational burden of the considered transceiver architecture is the following. $2M$ FFTs of length $k$ are required, with a complexity proportional to $2M k \log_2 k$; in order to compute the zero-forcing matrix, the FFT of the matrix-valued sequence $\widetilde{{\cal H}}(n)$ must be computed, with a complexity proportional to $MN_tT (k \log_2 k)$; computation of the matrix $(\widetilde{{\cal H}}(n) \mathbf{Q})$ and of its inverse, for $n=1, \ldots, k$,  finally requires a computational burden proportional to $k(N_T M^2 + M^3)$. 

It can be easily seen that the complexity of the FDE scheme is much lower than that of the TDE scheme.

\subsection{Waveform choice} \label{sec:3}

In this section, we describe some shaping pulses that are currently being considered as alternatives to the rectangular pulse adopted in OFDM and that can be used also as shaping transmit and receive filters in our considered modulation schemes.
In practice, we are interested in pulses that achieve a good compromise between their sidelobe levels in the frequency domain, and their extension in the time-domain. We report here three possible examples of pulse shapes, namely the evergreen root-raised cosine (RRC), the pulse proposed in the PHYDYAS research project \cite{PHYDIAS-D5.1-2008} for use with the Filterbank Multi-Carrier modulation, and, finally, the Dolph-Chebyshev (DC) pulse. 

\textbf{RRC pulses} are widely used in telecommunication systems to minimized ISI at the receiver. The impulse response of an RRC pulse is
\begin{equation}
\begin{array}{lll}
p(t) =  \\
\left\lbrace
\begin{array}{l l}
\frac{1}{\sqrt{T}}\left(1-\alpha+4\frac{\alpha}{\pi}\right) & t=0\\[3mm]
\frac{\alpha}{\sqrt{2T}}\left[\left(1+\frac{2}{\pi}\right)\sin\left(\frac{\pi}{4\alpha}\right)+\left(1-\frac{2}{\pi}\right)\cos\left(\frac{\pi}{4\alpha}\right)\right] & t=\pm\frac{T}{4\alpha}\\[3mm]
\frac{1}{\sqrt{T}}\frac{\sin\left(\pi\frac{t}{T}\left(1-\alpha\right)\right)+4\alpha\frac{t}{T}\cos\left(\pi\frac{t}{T}\left(1+\alpha\right)\right)}{\pi\frac{t}{T}\left[1-\left(4\alpha\frac{t}{T}\right)^2\right]} & {\rm otherwise}
\end{array}
\right.
\end{array}
\end{equation}
where $T$ is the symbol interval and $\alpha$ is the roll-off factor, which measures the excess bandwidth of the pulse in the frequency domain.

\textbf{The  PHYDYAS pulse} is a discrete-time pulse specifically designed for FBMC systems. Let $M_s$ be the number of subcarriers, then the impulse response is
$$
p(n)=P_0+2\sum_{k=1}^{K-1} (-1)^k P_k \cos\left(\frac{2\pi k}{KM}(n+1)\right)\,,
$$
for $n=0,1,\ldots,KM-2$ and $K=4$, where the coefficients $P_k,~k=0,\ldots,K-1$ have been selected using the frequency sampling technique~\cite{PHYDIAS-D5.1-2008}, and assume the following values:
\begin{eqnarray}
P_0&=&1 \nonumber\\
P_1&=&0.97195983 \nonumber\\
P_2&=&1/\sqrt{2} \nonumber\\
P_3&=&\sqrt{1-P_1}\,. \nonumber
\end{eqnarray}

 \textbf{The DC pulse}~\cite{Do46} is significant because, in the frequency domain, it minimizes the main lobe width for a given side lobe attenuation. Its discrete-time impulse response is~\cite{Antoniou2000}
$$
\begin{array}{lll}
p(n)=\frac{1}{N}\left[10^{-\frac{A}{20}}+2\sum_{k=1}^{(N-1)/2} T_{N-1}\left(x_0 \cos\left(\frac{k\pi}{N}\right)\right)\cos\left(\frac{2\pi nk}{N}\right)\right]\,,
\end{array}
$$
for $n=0,\pm 1,\ldots,\pm \frac{N-1}{2}$, where $N$ is the number of coefficients, $A$ is the attenuation of side lobes in dB, 
$$
x_0=\cosh\left(\frac{1}{N-1}\cosh^{-1}\left(10^{-\frac{A}{20}}\right)\right)\,,
$$
and
$$
T_n(x)=\left\lbrace
\begin{array}{l l}
\cos\left(n \cos^{-1}(x)\right) & |x|\le 1\\
\cosh\left(n \cosh^{-1}(x)\right) & |x|> 1\\
\end{array}
\right.
$$
is the Chebyshev polynomial of the first kind~\cite{AbSt72}.

\begin{figure}[t]
\centering
\includegraphics[scale=0.30]{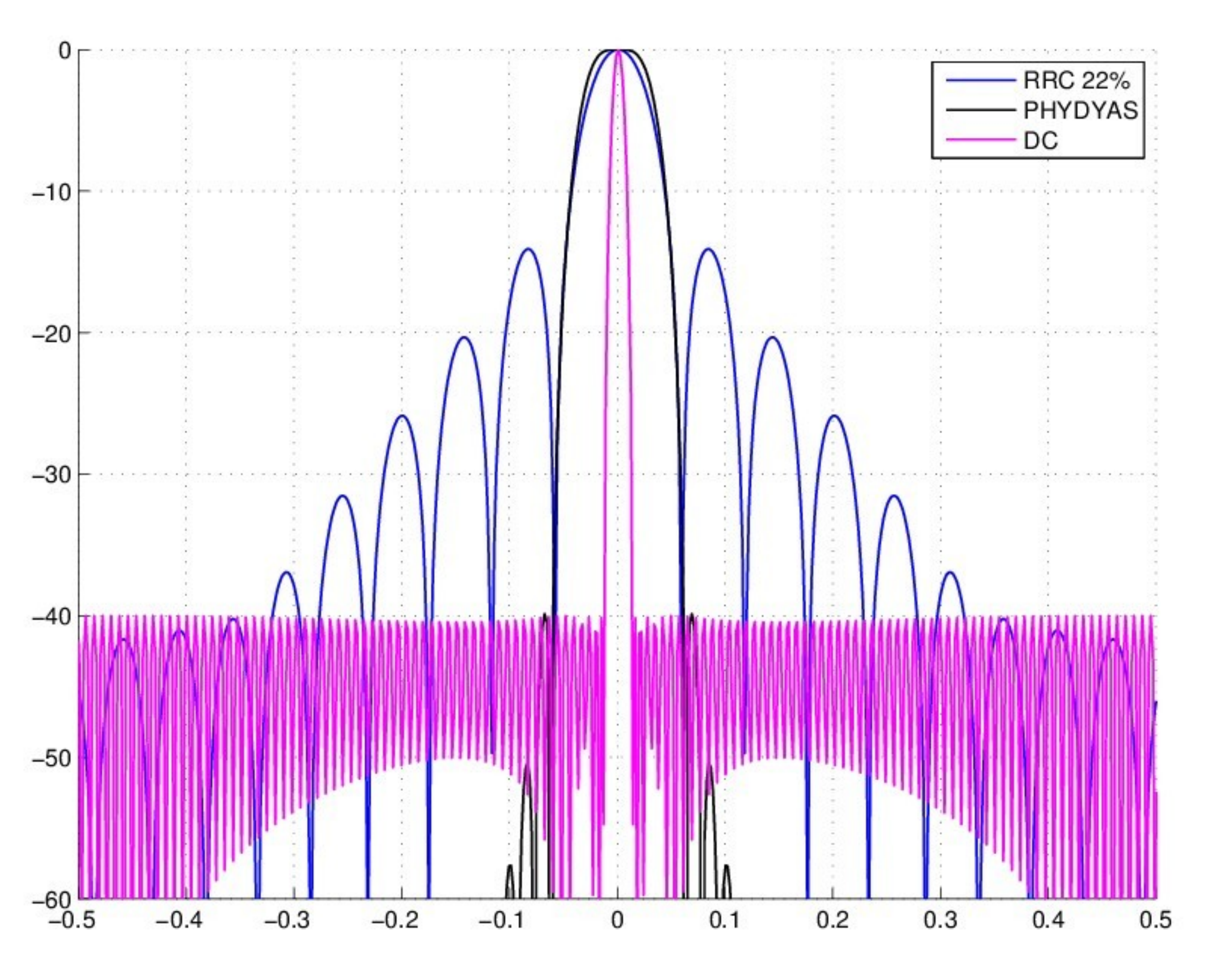}
\caption{Comparison of pulse shapes in the frequency domain.}
\label{Fig:spettri}
\end{figure}

In Fig.~\ref{Fig:spettri}, we report the spectra of the pulses we have just described. All spectra were computed by performing a 1024 points FFT of pulses of 256 samples in the time domain. The figure compares an RRC pulse having roll-off $\alpha=0.22$, the PHYDYAS pulse with $M_s=1$, and the DC pulse with attenuation $A=-50$ dB. The figure clearly shows that the rectangular pulse is the one with the worst spectral characteristics; on the other hand, the PHYDYAS pulse is the one with the smallest sidelobe levels, while the DC pulse is the one with the smallest width of the main lobe. 

\section{Computation of the achievable spectral efficiency}

As a figure of merit to compare the different transceiver architectures  with the different employed pulses, we will use the ASE, that is the maximum achievable spectral efficiency with the constraint of arbitrarily small BER.  The ASE takes the particular constellation and signaling parameters into consideration, so it does not qualify as a normalized capacity measure (it is often called \textit{constrained capacity}). We evaluate only ergodic rates so the {ASE} is computed given the channel realization  and averaged over it---remember that we are assuming perfect channel state information at the receiver. The spectral efficiency $\rho$ of any practical coded modulation system operating at a low packet error rate is upper bounded by the {ASE}, i.e., $\rho\leq \mathrm{ASE}$, where
\begin{equation}
\mathrm{ASE}=\frac{1}{T_{\rm s}W}\lim_{L\rightarrow\infty}\frac{1}{L} E_{\widetilde{\mathbf{H}}} \left[I(\mathbf{s};\hat{\mathbf{s}}|\widetilde{\mathbf{H}})\right]\;\mathrm{bit/s/Hz} \label{eq:ASE}
\end{equation}
$I(\mathbf{s};\hat{\mathbf{s}}|\tilde{\mathbf{H}})$ being the mutual information given the channel realization,  $T_{\rm s}$ the symbol interval, and $W$ the signal bandwidth (as specified in Section~\ref{sec:numerical_results}). Although not explicitly reported, for notational simplicity, the ASE in (\ref{eq:ASE}) depends on the SNR.

The computation of the mutual information requires the knowledge of the channel conditional probability density function (pdf) $p(\hat{\mathbf{s}}| \mathbf{s},\tilde{\mathbf{H}})$. It can be numerically computed by adopting the simulation-based technique described in~\cite{ArLoVoKaZe06} once the channel at hand is finite-memory and the optimal detector for it is available.
In addition, only the optimal detector for the actual channel is able to achieve the ASE in (\ref{eq:ASE}). 

In both transceiver models described in Section II the soft symbol estimates can be expressed in the form
\begin{equation}
\widehat{\mathbf{s}}(n)=\mathbf{A}\mathbf{s}(n)+ \sum_{\ell \neq 0} \mathbf{A}_{\ell} \mathbf{s}(n-\ell) + \mathbf{z}(n)
\label{eq:auxiliary}
\end{equation}
i.e., as a linear transformation (through matrix $\mathbf{A}$, which eventually is zero in the FDE case with zero-forcing equalization) of the desired QAM data symbols, plus a linear combination of the interfering data symbols and the colored noise $\mathbf{z}(n)$ having a proper covariance matrix. The optimal receiver has a computational complexity which is out of reach and for this reason we consider much simpler linear suboptimal receivers. Hence,  we are interested in the achievable performance when using suboptimal low-complexity detectors. We thus resort to the framework described in~\cite[Section VI]{ArLoVoKaZe06}. We compute proper lower bounds on the mutual information
 (and thus on the ASE) obtained by substituting $p(\hat{\mathbf{s}}| \mathbf{s},\tilde{\mathbf{H}})$ in the mutual information definition with an arbitrary auxiliary channel law $q(\hat{\mathbf{s}}| \mathbf{s},\tilde{\mathbf{H}})$
with the same input and output alphabets as the original channel (mismatched detection~\cite{ArLoVoKaZe06})---the more accurate the auxiliary channel to approximate the actual one, the closer the bound.
If the auxiliary channel law can be represented/described as a finite-state channel, the
pdfs $q(\hat{\mathbf{s}}|\mathbf{s}, \tilde{\mathbf{H}})$ and $q_p(\hat{\mathbf{s}}| \tilde{\mathbf{H}})=\sum_{\mathbf{s}}q(\hat{\mathbf{s}}|\mathbf{s}, \tilde{\mathbf{H}})P(\mathbf{s})$
can be computed, this time, by using the optimal maximum a posteriori symbol detector
for that auxiliary channel \cite{ArLoVoKaZe06}. This detector, that
is clearly suboptimal for the actual channel, has at its
input the sequence $\hat{\mathbf{s}}$ generated by simulation \emph{according
to the actual channel model} (for details, see \cite{ArLoVoKaZe06}).  If we
change the adopted receiver (or, equivalently, if we change the auxiliary
channel) we obtain different lower bounds on the constrained capacity
but, in any case, these bounds are \emph{achievable} by those receivers,
according to mismatched detection theory~\cite{ArLoVoKaZe06}.
We thus say, with a slight abuse of terminology, that the computed
lower bounds are the ASE values of the considered channel when those
receivers are employed.

This technique thus allows us to take reduced-complexity receivers into account. In fact, it is sufficient to consider an auxiliary channel which is a simplified version of the actual channel in the sense that
only a portion of the actual channel memory and/or a limited number of impairments are present. In particular, we will use the auxiliary channel law (\ref{eq:auxiliary}), where the sum of the interference and the thermal noise
$\mathbf{z}(n)$ is assimilated to Gaussian noise with a proper covariance matrix.


 
The transceiver models with the different shaping pulses are compared in terms of ASE without taking into account specific coding schemes, being understood that, with a properly designed channel code, the information-theoretic performance can be closely approached.

\begin{figure}[t]
\centering
\includegraphics[scale=0.27]{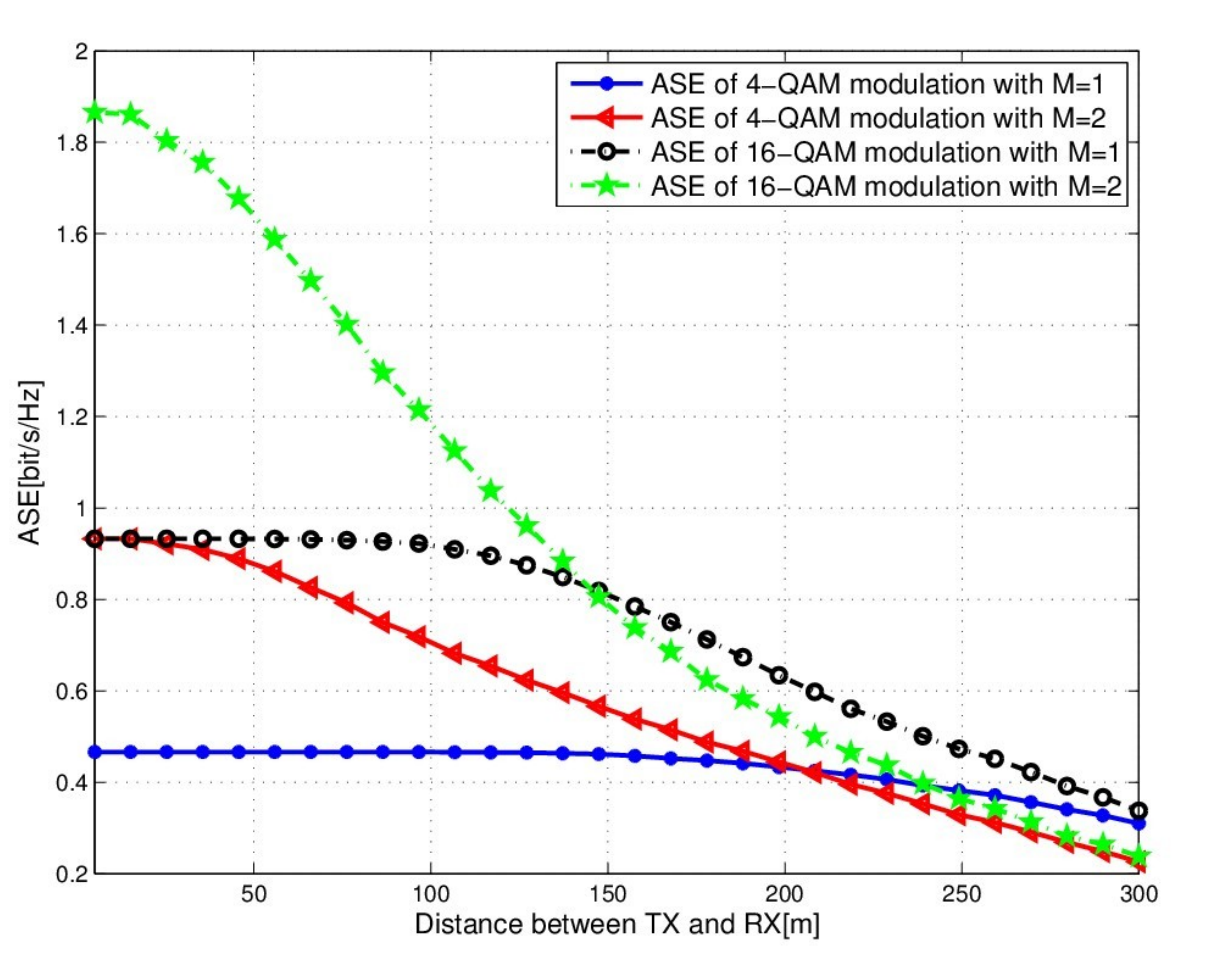}
\caption{ASE versus distance; impact of modulation cardinality and multiplexing order. Parameters: DC pulse; $P_T=0$dBW.}
\label{Fig:fig5}
\end{figure}

\begin{figure}[t]
\centering
\includegraphics[scale=0.27]{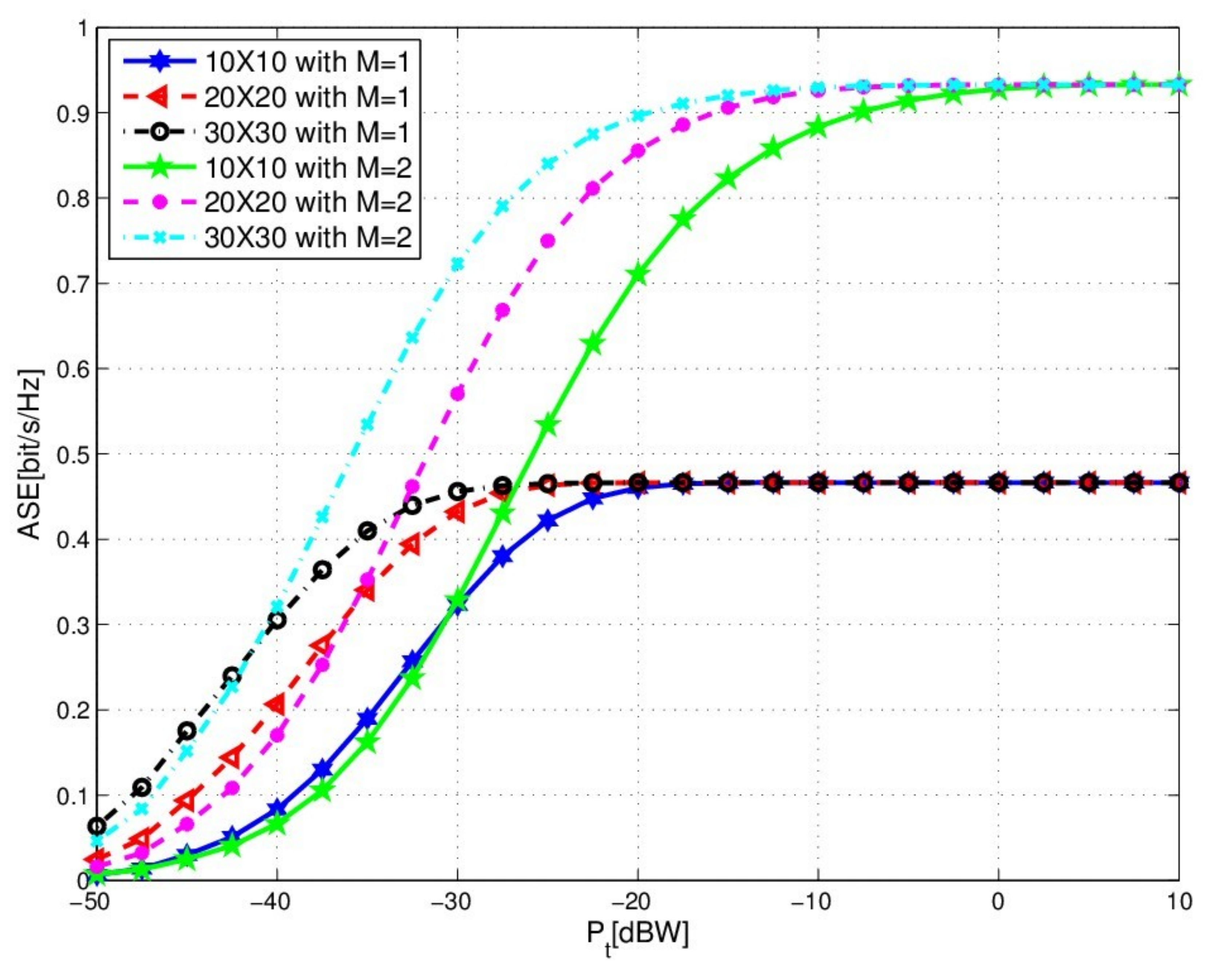}
\caption{ASE versus transmit power; impact of array size and multiplexing order. Parameters: 4-QAM modulation; DC pulse; $d=30$m; varying $N_R \times N_T$.}
\label{Fig:fig6}
\end{figure}

\begin{figure}[t]
\centering
\includegraphics[scale=0.27]{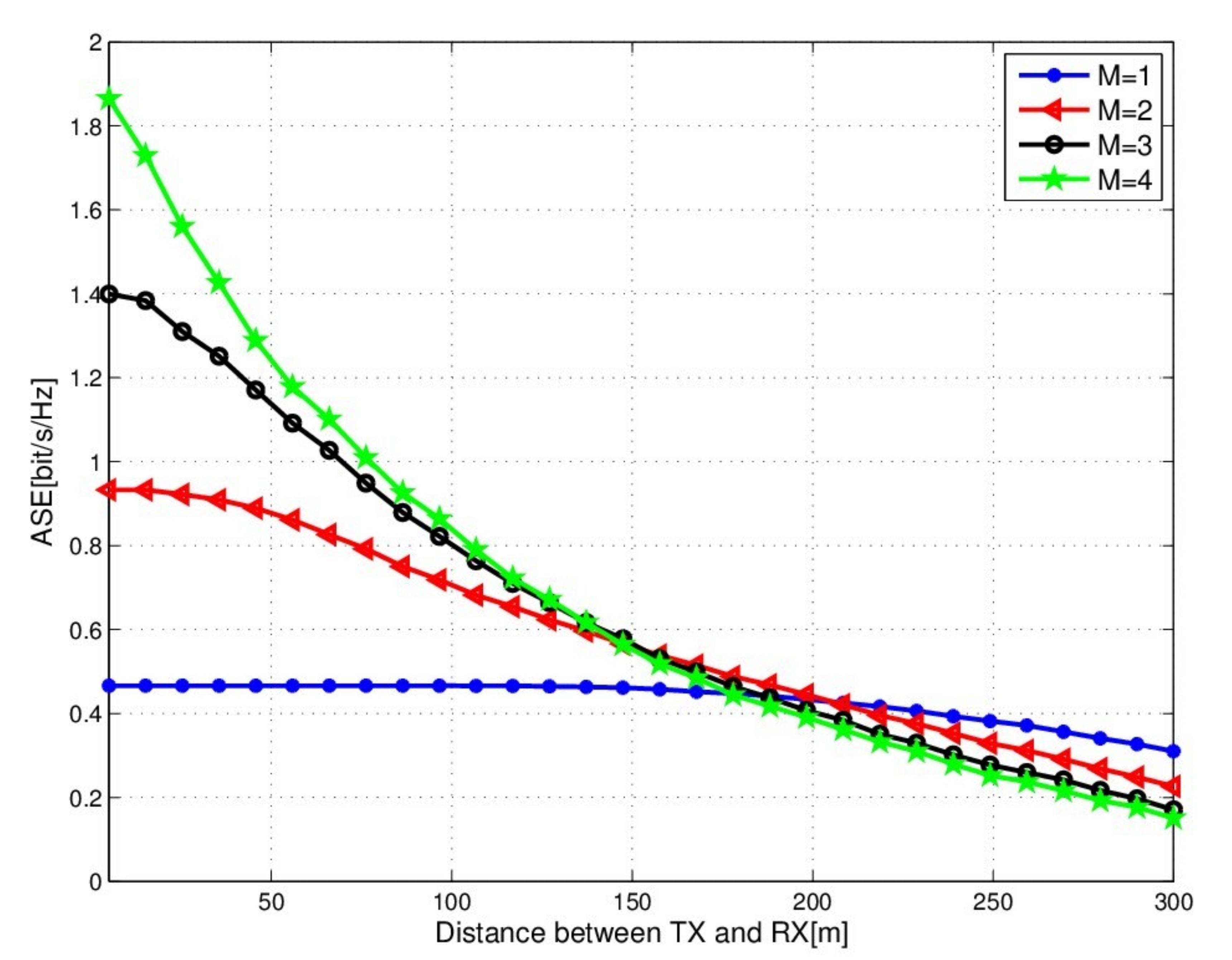}
\caption{ASE versus distance; impact of multiplexing order. Parameters: 4-QAM modulation; DC pulse; 
$P_T=0$dBW; $N_R \times N_T= 10 \times 10$.}
\label{Fig:fig7}
\end{figure}

\section{Numerical Results} \label{sec:numerical_results}
\begin{figure}[t]
\centering
\includegraphics[scale=0.27]{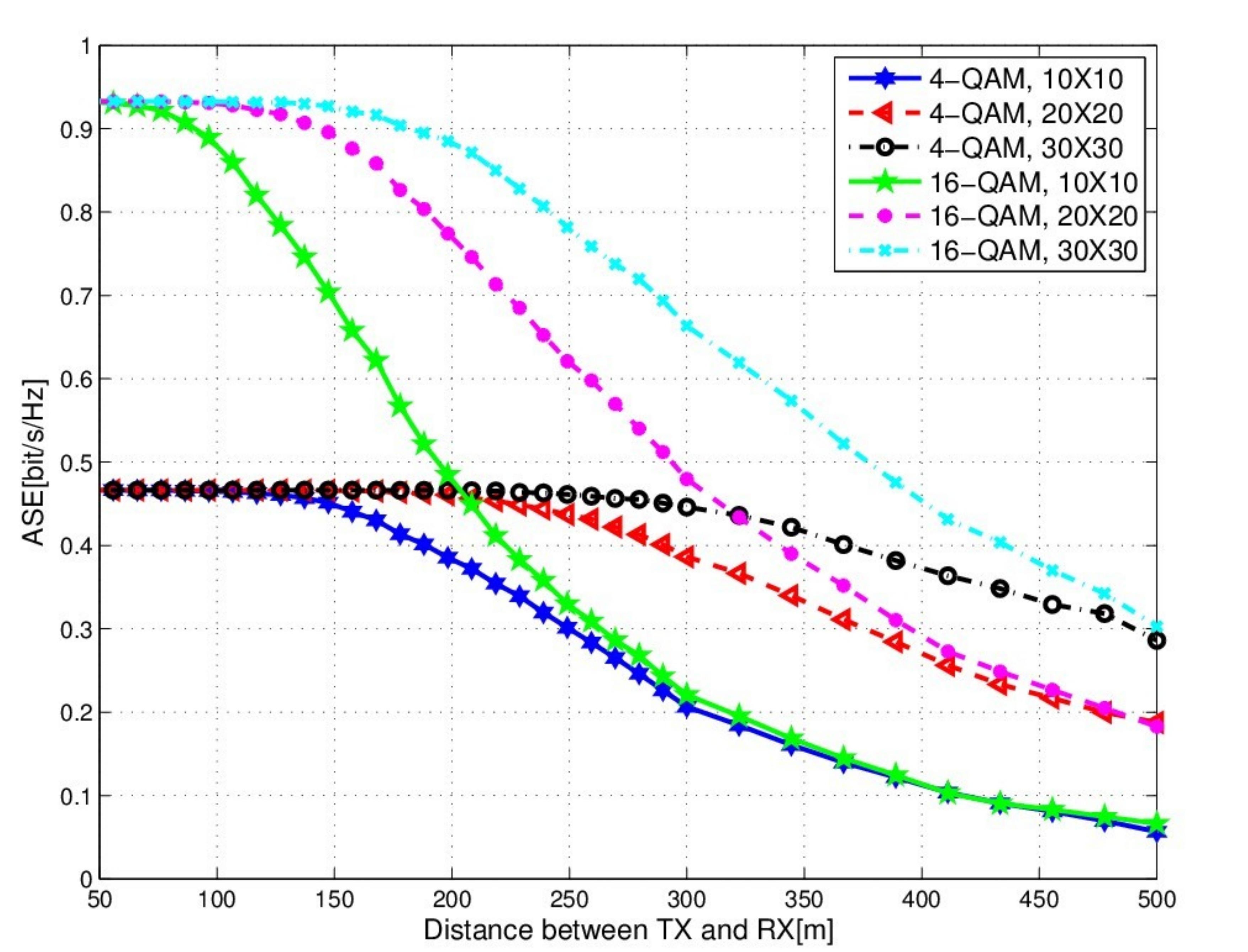}
\caption{ASE versus distance; impact of modulation cardinality and array size.
Parameters: DC pulse; $P_T=3$dBW; $M=1$; verying $N_R \times N_T$.}
\label{Fig:fig9}
\end{figure}

We now report some simulation results.
We consider a communication bandwidth of $W=500$MHz centered over a mmWave carrier frequency. The MIMO propagation channel has been generated according to the statistical procedure detailed in \cite{spatiallysparse_heath,heath_selective}, with a path-loss exponent equal to 3.3 \cite{tractable}. The additive thermal noise is assumed to have a power spectral density of -174dBm/Hz, while the front-end receiver is assumed to have a noise figure of 3dB. We study, in the following figures, the ASE for varying values of the transmit power $P_t$, of the distance $d$ between the transmitter and the receiver, of the number of transmit and receive antennas, of the multiplexing order $M$, and for the case in which  the PHYDYAS pulse is adopted\footnote{A deeper analysis about the impact of the choice of different pulses will form the object of future research.}. 
For this waveform, we define the bandwidth as the frequency range such that out-of-band emissions are 40dB below the maximum in-band value of the Fourier transform of the pulse.  For the considered communication bandwidth of $W=500$MHz, we found that the symbol interval $T_{\rm s}$ is 3.96ns for the PHYDYAS pulse, for the case in which we consider its truncated version to the interval $[-4T_{\rm s}, 4T_{\rm s}]$.
The reported results are to be considered as an ideal benchmark for the ASE since we are neglecting the interference\footnote{We note however that being mmWave systems mainly noise-limited rather than interference limited, the impact of this assumption on the obtained results is very limited.}, and we are considering digital pre-coding and post-coding, whereas due to hardware constraints mmWave systems will likely operate with hybrid analog/digital beamforming strategies \cite{hybridprecodingmmwaves}\footnote{The evaluation of the ASE with hybrid analog/digital pre-coding and post-coding structures is an interesting issue that is out of the scope of this paper but certainly worth future investigation.}. We focus here on the performance of the TDE transceiver, since our tests showed that the FDE structure is worse than the TDE scheme.
Fig.s \ref{Fig:fig5}, \ref{Fig:fig7} and \ref{Fig:fig9} report the ASE\footnote{Of course, the achievable rates in bit/s can be immediately obtained by multiplying the ASE by the communication bandwidth $W=500$MHz.} versus the distance $d$ between the transmitter and the receiver, assuming that the transmit power is $P_t=0$dBW, while Fig. \ref{Fig:fig6}  reports the ASE versus the transmit power $P_t$ (varying in the range $[-50, 10]$dBW), assuming a link length $d=30$ m. 
Inspecting the figures, the following remarks are in order:
\begin{itemize}
\item[-]
Results, in general, improve for increasing transmit power, for decreasing distance $d$ between transmitter and receiver
and for increasing values of the number of transmit and receive antennas. 
\item[-]
In particular, good performance can be attained for distances up to 100 - 200m, whereas for $d>200m$ we have 
a steep degradation of the ASE. In this region, all the advantages given by increasing the modulation cardinality or the number of antennas are essentially lost or reduced at very small values. Of course, this performance degradation may be compensated by increasing the transmit power. 
\item[-]
Regarding the multiplexing index $M$, it is interesting to note from Fig. \ref{Fig:fig7} that for short distances the system benefits from a large multiplexing order, while, for large distances (which essentialy corresponds to low signal-to-noise ratio), the ASE corresponding to $M=1$ is larger than that corresponding to the choise $M>1$.
\item[-]
For a reference distance of 30m (which will be a typical one in small-cell 5G deployments for densely crowded areas), a trasnmit power around 0dBW is enough to grant good performance and to benefit from the advantages of increased modulation cardinality, size of the antenna array, and multiplexing order.
\end{itemize}


\section{Conclusion}
This paper has provided a preliminary assessment of the ASE for a MIMO link operating at mmWave frequencies with SCM. Two different transceiver architectures have been considered, one with time-domain equalization and one with cyclic prefix plus frequency domain equalization. Results have been shown with reference to the  TDE structure, which was found to outperform the FDE structure. For distances up to 100m and for values of the transmit power around 0dBW  a good performance level can be attained, with ASE values up to 1.8 bit/s/Hz, which, for a bandwidth of 500MHz, leads to a bit-rate of up to almost 1Gbit/s.
The present study can be generalized and strengthened  in many directions. First of all, the impact of hybrid analog/digital beamforming should be evaluated; moreover, the considered analysis might be applied to a point-to-multipoint link, wherein the presence of multiple antennas at the transmitter is used for simultaneous communication with distinct users (the so-called multiuser MIMO technique). Additionally, since, as already discussed, the reduced wavelength of mmWave permits installing arrays with many antennas in small volumes, an analysis, possibly through asymptotic analytic considerations, of the very large number of antennas regime could also be made. Last, but not least, energy-efficiency considerations should also be made: both the ASE and the transceiver power consumption  increase for increasing transmit power and increasing size of the antenna arrays; if we focus on the ratio between the ASE and the transceiver power consumption, namely on the system energy efficiency, optimal trade-off values for the transmit power and size of the antenna arrays should be found. These topics are certainly worth future investigation.





%
\bibliographystyle{IEEEtran}

\end{document}